\documentclass[traditabstract]{aa}

\usepackage{graphicx}
\usepackage{multirow}
\usepackage{natbib}
\usepackage{adjustbox}
\usepackage{txfonts}
\usepackage{lipsum} 
\usepackage{mathrsfs}
\usepackage{breqn}
\usepackage[colorlinks = true,linkcolor = blue,urlcolor = blue,citecolor = blue]{hyperref}
\usepackage{xcolor}
\definecolor{yellowgray}{rgb}{0.90, 0.90, 0.2}
\definecolor{bluegray}{rgb}{0.20, 0.60, 0.80}
\definecolor{palered}{rgb}{0.99, 0.40, 0.5}
\definecolor{darkgray}{rgb}{0.35, 0.35, 0.35}
\definecolor{darkgrayb}{rgb}{0.75, 0.75, 0.75}
\definecolor{palegray}{rgb}{0.96, 0.96, 0.96}

\begin{document} 

\title{Non-Local Thermodynamic Equilibrium inversions of the \ion{Si}{i} 10827~\AA \ spectral line}
  \titlerunning{NLTE inversions of the \ion{Si}{i} 10827~\AA \ transition}
   \author{C. Quintero Noda\inst{1,2} \and N. G. Shchukina\inst{3} \and A. Asensio Ramos\inst{1,2} \and M. J. Martínez González\inst{1,2} \and T. del Pino Alemán\inst{1,2} \and \\ J. C. Trelles Arjona\inst{1,2} \and M. Collados\inst{1,2}
             }
   \institute{Instituto de Astrof\'isica de Canarias, E-38205, La Laguna, Tenerife, Spain.   \\
     \email{carlos.quintero@iac.es}    
\and  Departamento de Astrof\'isica, Univ. de La Laguna, La Laguna, Tenerife, E-38200, Spain
\and Main Astronomical Observatory, National Academy of Sciences, 03143 Kyiv, Ukraine}
   \date{Received 2024/09/13 ; accepted 2024/11/16  }

\abstract{Inferring the coupling of different atmospheric layers requires observing spectral lines sensitive to the atmospheric parameters, particularly the magnetic field vector, at various heights. The best way to tackle this goal is to perform multi-line observations simultaneously. For instance, the new version of the Gregor Infrared Spectrograph instrument offers the possibility to observe the spectral lines at 8542 and 10830~\AA \ simultaneously for the first time. The first spectral window contains the \ion{Ca}{ii} 8542~\AA \ spectral line, while the \ion{Si}{i} 10827~\AA \ transition and \ion{He}{i} 10830~\AA \ triplet infrared lines can be found in the second spectral window. As the sensitivity to the atmospheric parameters and the height of formation of those transitions is different, combining them can help understand the properties of the solar photosphere and chromosphere and how they are magnetically coupled. Traditionally, the analysis of the \ion{Si}{i} 10827~\AA \ transition assumes local thermodynamic equilibrium (LTE), which is not the best approximation to model this transition. Hence, in this work, we examine the potential of performing non-LTE (NLTE) inversions of the full Stokes vector of the \ion{Si}{i} 10827~\AA \ spectral line. The results indicate that we properly infer the atmospheric parameters through an extended range of atmospheric layers in comparison with the LTE case (only valid for the spectral line wings, i.e., the low photosphere), with no impact on the robustness of the solution and just a minor increase in computational time. Thus, the NLTE assumption will help to accurately constrain the photospheric physical parameters when performing combined inversions with, e.g., the \ion{Ca}{ii} 8542~\AA \ spectral line.}

\keywords{Sun: magnetic fields, chromosphere -- Techniques: polarimetric, high angular resolution --  Radiative transfer}

\maketitle

\section{Introduction}

Understanding solar phenomena requires access to the physical parameters over a particular spatial domain and a specific range of atmospheric heights. The reason is that solar phenomena, particularly their magnetic field, can extend vertically (with respect to the solar surface) several hundred kilometres. Hence, to understand how the magnetic field is connected and structured along different atmospheric layers, we need to analyse spectral lines sensitive to the atmospheric parameters over such a range of heights. The best strategy is to simultaneously analyse spectral lines with different sensitivities to the atmospheric parameters and height of formation. In this regard, multi-line spectropolarimetric observations are the most common strategy to fulfil the mentioned requirement. The large-scale solar telescopes Daniel K. Inouye Solar Telescope \citep{Rimmele2020} and the future European Solar Telescope \citep{QuinteroNoda2022EST} have been designed to offer these multi-line configurations. In addition, instruments like the Gregor Infrared Spectrograph \citep[GRIS,][]{Collados2012} installed at the Gregor telescope \citep{Schmidt2012} have been recently upgraded to support multi-channel observations of complementary spectral lines \citep{QuinteroNoda2022GRIS}. In particular, GRIS was upgraded to observe simultaneously \ion{Ca}{ii} 8542~\AA, with high sensitivity to the atmospheric parameters in the lower and middle chromosphere \citep[see, for instance,][]{QuinteroNoda2016}, and the \ion{He}{i} 10830~\AA \ infrared triplet whose formation properties make them ideal to examine targets such as filaments, spicules, and prominences.

Additionally, the spectral window at 10830~\AA \ also contains the \ion{Si}{i} 10827~\AA \ transition. This strong photospheric line can complement the other two transitions when inferring the properties of the magnetic field at photospheric layers. However, although an accurate determination of the atomic populations of the silicon transition requires solving the radiative transfer equation under non-Local Thermodynamic Equilibrium \citep[NLTE, e.g.,][]{Bard2008,2012KPCB...28..169S,Shchukina2017,2019A&A...628A..47S} the spectral line has usually been treated under the Local Thermodynamic Equilibrium (LTE) formalism \citep[with some exceptions such as in][]{2017A&A...607A.102O} when performing inversions \citep[see, for instance,][]{2012A&A...542A.112K,2018A&A...617A..39F,2019A&A...625A.128D}. The main reason for this is that certain atoms, such as iron, calcium, or silicon, require a large number of atomic levels and transitions for their accurate modelling (particularly in the case of the minority species of the atom), which results in atomic models that are computationally expensive to work with. In order to reduce the computational burden, hybrid methods have been used in the past. For instance, \cite{2012A&A...539A.131K} used NLTE departure coefficients computed with the atomic model presented in \cite{Bard2008}. In this case, the authors used a set of fixed departure coefficients to study the differences between a pure LTE inversion and one corrected by the mentioned departure coefficients. Their findings revealed no noticeable differences between LTE and NLTE for the magnetic field inclination and azimuth, while more significant variations were detected for the magnetic field strength. Moreover, the authors explained that even more prominent differences between LTE and NLTE were found for the inferred temperature.

In preparation for the observations that GRIS has already started \citep[][]{QuinteroNoda2024First} and those soon to be provided by the Diffraction Limited Near-Infrared Spectropolarimeter \citep[DL-NIRSP;][]{2022SoPh..297..137J} we aim to study the differences between LTE versus NLTE inversions with the simplified model atom presented in \cite{Shchukina2017}. The goal is to analyse whether there is an improvement in the inferred atmosphere, particularly the thermal atmospheric parameters when using such a simplified atom, quantify the increase in computational time, and evaluate the robustness of the inversion process to use it as a baseline for the analysis of future observations from GRIS and similar instruments.

\section{Data and methodology}\label{Method}

The main goal of this work is to examine the difference between performing inversions of synthetic spectra of the \ion{Si}{i} 10827~\AA \ transition under LTE and NLTE conditions. We use the Stokes Inversion code based on Response functions \citep[SIR,][]{RuizCobo1992} for the LTE inversions and the Departure coefficient aided Stokes Inversion based on Response functions \citep[DeSIRe;][]{DeSIRe} code for the NLTE inversions. The DeSIRe code can work both under the LTE and the NLTE approximation, but, as the reference inversions that have been done in the past for the \ion{Si}{i} 10827~\AA \ transition were done mainly with the SIR code, we believe it is appropriate to compare the inversion results from SIR with those from DeSIRe assuming NLTE.

To benchmark these inversions, we use two types of model atmospheres. We employ the semi-empirical atmosphere Harvard-Smithsonian Reference Atmosphere \citep[HSRA;][]{1971SoPh...18..347G} for comparing the main differences between the intensity profiles in LTE, NLTE, and the solar atlas. We also use the HSRA atmosphere for estimating the response functions \citep[RF; see, e.g.,][]{1977A&A....56..111L} to perturbations of the atmospheric parameters in LTE and NLTE. We also employ a 3D realistic magneto-hydrodynamic numerical simulation, specifically snapshot 385 of the enhanced network simulation\footnote{\url{http://sdc.uio.no/search/simulations}} described in \cite{2016A&A...585A...4C} and carried out with the Bifrost code \citep{2011A&A...531A.154G}. The snapshot covers a field-of-view (FOV) of $24\times24$~Mm$^2$ with a pixel size of 48~km, while the vertical domain extends from 2.4~Mm below to 14.4~Mm above the average height at which optical depth is unity at $\lambda=5000$~\AA. 

We assume disk centre observations, that is, $\mu=\cos(\theta)=1$, where $\theta$ is the heliocentric angle and consider the abundance values given in \cite{Asplund2009}. We use a spectral sampling of 20~m\AA \ and degrade the synthetic Stokes profiles by adding noise with an amplitude of $5\times10^{-4}$ of $I_c$. We preserve the original spatial resolution of 48 km per pixel, close to the value at the diffraction limit that Gregor could achieve at 10827~\AA.

Regarding the atomic model for solving the atomic populations of the silicon transition, we use the simplest model presented in \cite{Shchukina2017}. This simplest working model has 6 bound-bound radiative transitions between fine structure levels of the 4s $^{3}$P$^{\rm o}$ (lower) and 4p $^{3}$P (upper) terms of \ion{Si}{i}. One of these transitions, namely 4s $^{3}$P$^{\rm o}_2$ - 4p $^{3}$P$_2$, produces the transition of interest \ion{Si}{i} 10827~\AA. The model also includes the levels that play an important role in the ionization balance. They are low-excitation meta-stable levels of \ion{Si}{i} lying below the 4s $^{3}$P$^{\rm o}$ term, including their corresponding parent levels in the next ionization stage of \ion{Si}{ii}. In total, the simplest model has 16 levels, with twelve atomic energy levels of \ion{Si}{i}, three levels from the 3p $^{2}$P$^{\rm o}$ and 3p $^{4}$P terms of \ion{Si}{ii}, the ground level 3s $^{1}$S$_1$ of \ion{Si}{iii}, that produce fifteen bound-free radiative transitions. We adapted the atom to be used with the RH code \citep{2001ApJ...557..389U,2003ApJ...592.1225U} and made it publicly available\footnote{\url{https://cloud.iac.es/index.php/s/caSrtC9KmQ8nWCN}} so it can be used by any modern inversion code such as those presented in \cite{2019A&A...623A..74D}, and \cite{2016ApJ...830L..24D} and \cite{2022ApJ...933..145L}.

\section{Synthetic profiles}\label{Results}

\begin{figure}
	\begin{center} 
		\includegraphics[trim=0 0 0 0,width=8.5cm]{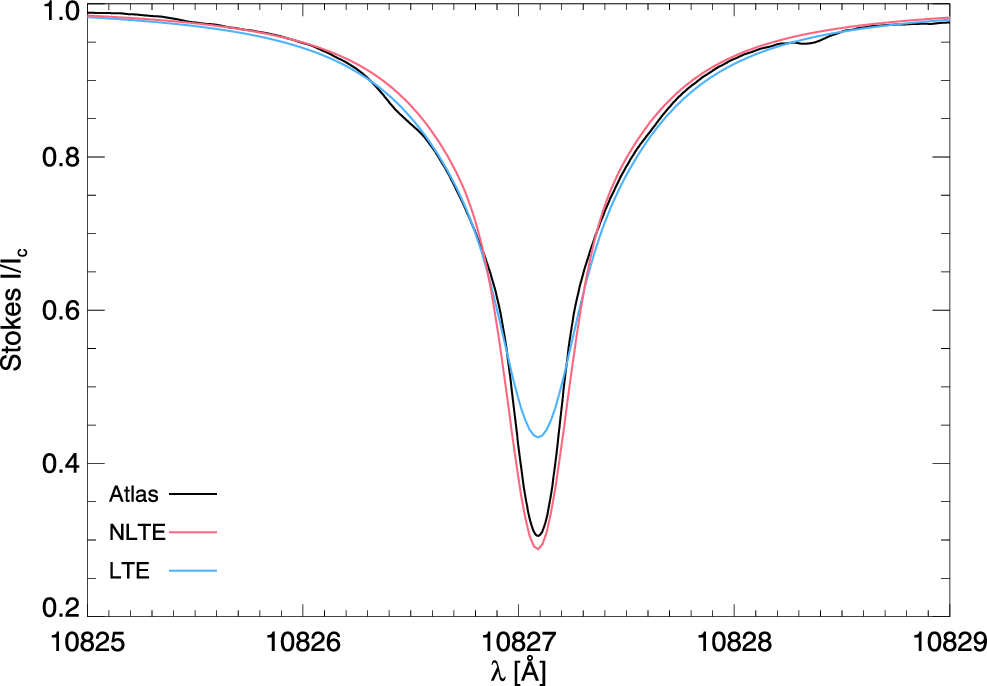}
		\vspace{-0.10cm}
		\caption{Intensity profiles for the \ion{Si}{i} 10827~\AA \ transition from the solar atlas (black), LTE synthesis (blue), and an NLTE synthesis (red). Both syntheses were computed using the HSRA model atmosphere.}
		\label{Atlas}
	\end{center}
\end{figure}

We examine the differences between solving the populations for the \ion{Si}{i} 10827~\AA \ transition in LTE versus NLTE. In particular, we make use of the HSRA atmosphere for the synthesis, and we compare the results with the atlas observed by \cite{Delbouille1973} and downloaded from the BAse de données Solaire Sol (BASS2000) archive\footnote{\url{http://bass2000.obspm.fr/solar_spect.php}}. Figure~\ref{Atlas} shows the comparison between the three intensity profiles. The LTE profile generated with SIR (blue) matches the continuum and the wings of the line with high accuracy while it is far from the atlas at line core wavelengths. This was expected, as the spectral line is sensitive to NLTE effects \citep[e.g.,][]{2012KPCB...28..169S,Shchukina2017,2019A&A...628A..47S}. The NLTE intensity profile (red) generated with DeSIRe is very close to the atlas for all wavelengths, which is indicative of the suitability of the simplified atom for modelling the silicon transition.

\begin{figure*}
	\begin{center} 
		\includegraphics[trim=0 0 0 0,width=18.0cm]{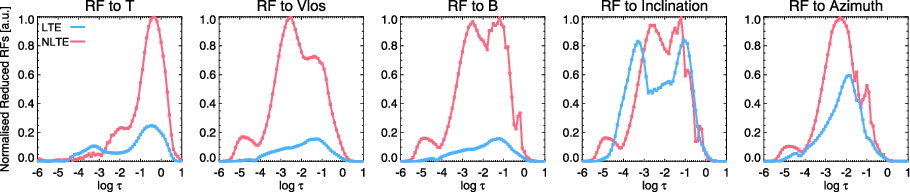}
		\vspace{-0.10cm}
		\caption{From left to right, RFs to changes in temperature, line-of-sight velocity, and the three components of the magnetic field vector in polar coordinates. Each curve corresponds to the maximum value for the four Stokes parameters and all the computed wavelengths at each optical depth for the \ion{Si}{i} 10827~\AA \ spectral line. Blue and red designate the results from the LTE and NLTE solutions, respectively. RFs are normalised to the maximum value of the corresponding NLTE solution.}
		\label{RFs}
	\end{center}
\end{figure*}

\begin{figure*}
	\begin{center} 
		\includegraphics[trim=0 0 00 0,width=18cm]{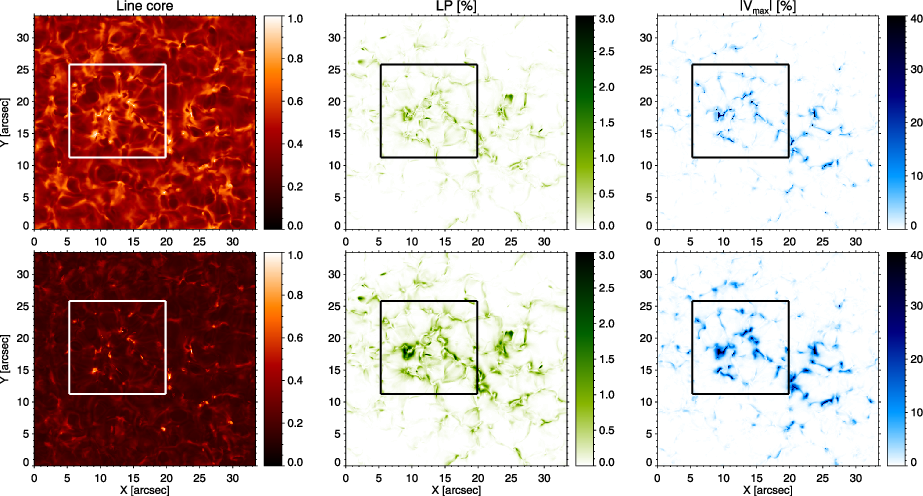}
		\vspace{-0.00cm}
		\caption{From left to right, we show the spatial distribution of line core intensity, maximum linear polarisation, and maximum circular polarisation signals for the \ion{Si}{i} 10827~\AA \ spectral line. The top row corresponds to the results from the LTE computation, while the bottom panels display the results of the NLTE case. The region enclosed by the white/black square corresponds to the FOV used for the inversion analysis in Section~\ref{Sec_inver}.}
		\label{Spatial}
	\end{center}
\end{figure*}

We also analyse the differences between the LTE and NLTE RFs due to changes in different atmospheric parameters. Using the HSRA atmosphere, we compute the RFs analytically with SIR and numerically with DeSIRe following the method explained in \cite{QuinteroNoda2016}. The results are presented in Figure~\ref{RFs}, where we show the maximum value of the RFs for all the wavelengths and the four Stokes parameters for the two approximations. Each RF is normalised to the maximum of the NLTE solution.

The RFs to changes in temperature in NLTE and LTE are almost identical in the lower layers of the model. However, they deviate at higher layers, with the NLTE RF dropping its maximum value just before $\log \tau=-3$. This result indicates that the sensitivity of the \ion{Si}{i} 10827~\AA \ spectral line to temperature changes is only significant at the lower and middle photosphere. These results seem similar to those found for other spectral lines like \ion{Na}{i} $D_1$ and $D_2$ \citep[e.g.,][]{Uitenbroek2006}, \ion{K}{i} $D_1$ and $D_2$ \citep[e.g., ][]{2017MNRAS.470.1453Q,2022A&A...666A.178A}, and the \ion{Mg}{i} b transitions \citep[e.g.,][]{2018MNRAS.481.5675Q}. Interestingly, the opposite behaviour is shown for the rest of the RFs, with the NLTE RFs having non-negligible contributions at higher layers than those in LTE. These results indicate that we can expect that inverting the \ion{Si}{i} 10827~\AA \ spectral line in NLTE will provide better accuracy for the inferred parameters at higher layers.

In order to expand the previous studies that used 1D semi-empirical atmospheres, from now on, we use the enhanced network 3D simulation for studying the impact of NLTE effects on the synthetic Stokes parameters. We first start synthesising the Stokes vector for the complete FOV of the simulation run. We show in Figure~\ref{Spatial} the spatial distribution of the line core intensity signals, maximum linear polarisation signals (computed as $LP = \sqrt{Q^2+U^2}$) and maximum circular polarisation signals in absolute value. The spatial distribution of intensity signals (left column) shows a higher contrast for the NLTE computation (bottom row). This agrees with the results presented in Figure~\ref{Atlas}, where the line core intensity in NLTE is deeper than the LTE line core. The spatial distribution of polarisation signals (middle and rightmost column) shows larger amplitudes in the magnetised regions for the entire FOV. In addition, there is no apparent deviation between the two computations besides the mentioned ones, i.e., the solar features seem to have the same shape and spatial distribution in both cases. 

\section{Spectropolarimetric inversions}\label{Sec_inver}

The main goal of this work is to ascertain whether we can improve the results obtained from LTE inversions of the \ion{Si}{i} 10827~\AA \ spectral line when using a simplified silicon atom to perform NLTE inversions. Additionally, two key elements are also essential. Firstly, we aim for an inversion process that is fast enough to be comparable to the LTE inversions. The NLTE inversion will always be more time-consuming, but we aim for a small time difference. Secondly, we want the inversion process to be as robust as possible. It is well-known that NLTE inversions are more unstable than LTE inversions, so we want to check the fraction of pixels in which the inversion code did not reach an accurate convergence.

To tackle those questions, we perform inversions of all the pixels enclosed in the squared area highlighted in Figure~\ref{Spatial}. We select this area because it is large enough to cover all the physical scenarios presented in the simulation, e.g., weakly magnetised areas and strongly magnetised ones. We use the synthetic profiles generated by DeSIRe and degraded with a noise contribution of $5\times10^{-4}$ of $I_c$. These will be the target profiles we fit. As we are able to obtain good fits in both LTE and NLTE inversions, our accuracy estimation will be directly how close the inferred atmospheres are to those used during the synthesis.

\begin{center}
	\begin{table}
		\vspace{+0.4cm}
		\caption{Number of nodes used for the inversion of each atmospheric parameter.} 	
		\begin{adjustbox}{width=0.45\textwidth}
			\bgroup
			\def\arraystretch{1.25}
			\begin{tabular}{|l|cccc|}
				\hline
				\multirow{2}{*}{\textbf{Parameter}}     & 
				\multicolumn{4}{c|}{\textbf{Nodes}}   \\
				& \textbf{Cycle 1} & \textbf{Cycle 2} &  \textbf{Cycle 3} & \textbf{Cycle 4}  \\
				\hline
				Temperature & 2 & 3 & 5 & 7 \\
				Line-of-Sight Velocity & 1 & 2 & 3 & 5 \\
				Microturbulence & 1 & 1 & 1 & 1 \\    
				Field Strength & 1 & 2 & 3 & 4 \\
				Inclination & 1 & 2 & 2 & 3 \\
				Azimuth & 1 & 1 & 1 & 2 \\
				\hline                       
			\end{tabular}
			\egroup
		\end{adjustbox}
		\label{tab:Config1}    
	\end{table}
\end{center} 

\begin{figure*}
	\begin{center} 
		\includegraphics[trim=0 0 00 0,width=18cm]{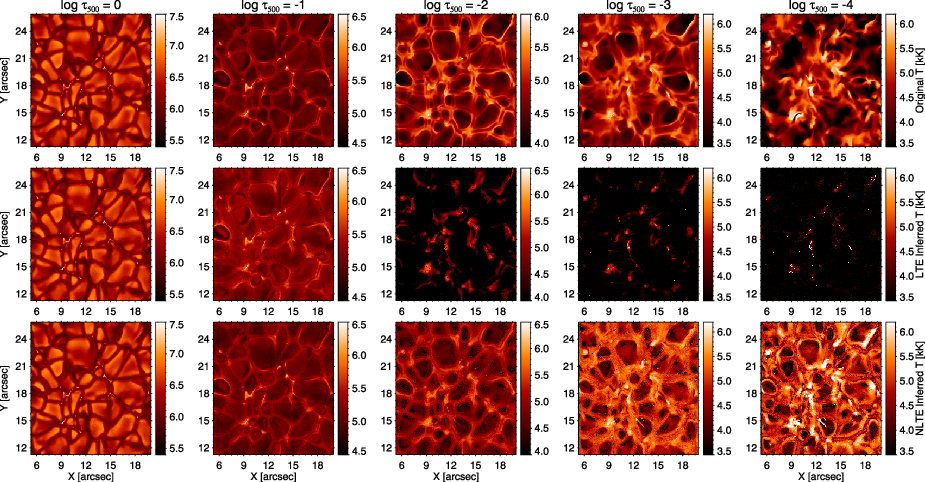}
		\vspace{-0.10cm}
		\caption{Spatial variation of the temperature at different optical depths (columns from left to right). Rows display, from top to bottom, the original atmosphere, the one inferred assuming LTE and that from NLTE inversions, respectively. The spatial domain corresponds to that highlighted by the square in Figure~\ref{Spatial}.}
		\label{Diff_T}
	\end{center}
\end{figure*}

Table~\ref{tab:Config1} shows the inversion configuration for the nodes. Each new column towards the rightmost side of the table adds (or maintains) more nodes for the inversion and corresponds to an additional inversion cycle. In addition, more information about the inversion strategy can be found in, e.g., \cite{Ferrente2024}. As a reminder, the inversion code uses as free parameters the perturbations of the physical quantities on a specific grid of optical depth points (nodes), taking into account that the inferred atmosphere will be the result of a high-order interpolation between those nodes (except in the case of 1 or 2 nodes, for which the interpolation is constant or linear, respectively). These free parameters are tuned to minimise the $\chi^2$ metric.

Regarding the physical quantities, we infer the temperature, the line-of-sight (LOS) velocity, microturbulence, and the magnetic field vector. The inversion is done in multiple cycles, increasing the number of nodes for each atmospheric parameter as presented in Table~\ref{tab:Config1}, except for the microturbulence, which is always set to a constant. As a note, although no microturbulent velocity was used during the synthesis because it comes from a 3D MHD simulation that did not have any, we still add it as a free parameter in the inversions to account for the complicated vertical stratification in the LOS velocity and improve the accuracy of the inversion results \citep[see the explanation in Sec.~3 of][for more information]{2019A&A...630A.133M}.

The inversion is done in parallel using the Python-based wrapper described in \cite{Gafeira2021}. This code allows the seamless utilisation of all available CPU cores to accelerate the inversion of many pixels. In addition, the wrapper can provide a set of atmospheres to initialise the inversion process, which allows initialising the inversion several times from different initial models and then picking the solution that achieves the best fit. This strategy helps in finding accurate solutions given the complexity of the $\chi^2$ hypersurface. Regarding the specific atmospheres used in the inversion as initialisation, they are created from the atmospheric models presented in \cite{1993fontenla} and the HSRA atmosphere, adding small amplitude variations in the LOS velocity, microturbulence, and magnetic field (which are null in the original models). We created a library of 12 initial atmospheres with constant values with height for those physical parameters to initialise the inversion of each pixel.

As a first step, we examine the spatial distribution of the atmospheric parameters at selected atmospheric layers. The goal is to compare the main differences between the input atmosphere and those inferred, assuming LTE and NLTE, respectively. Figure~\ref{Diff_T} shows the results for the temperature. The spatial distribution of the inferred atmospheres is very close to the input simulation at $\log \tau=0$ for both approximations. However, there are differences between the LTE and NLTE inversions in the middle and upper layers. At $\log \tau=-1$, the LTE inversion lacks contrast, while the NLTE inversion better resembles the input atmosphere. However, higher up, the LTE inversion is entirely different, indicating that the inferred atmospheres assuming LTE would only be accurate in the lower photosphere. In the case of NLTE inversions, there is a certain resemblance up to $\log \tau=-3$. However, at this height, the NLTE inversions cannot accurately reproduce the coolest regions, which translates into a lack of contrast with respect to the input atmosphere.

\begin{figure*}
	\begin{center} 
		\includegraphics[trim=0 0 00 0,width=18cm]{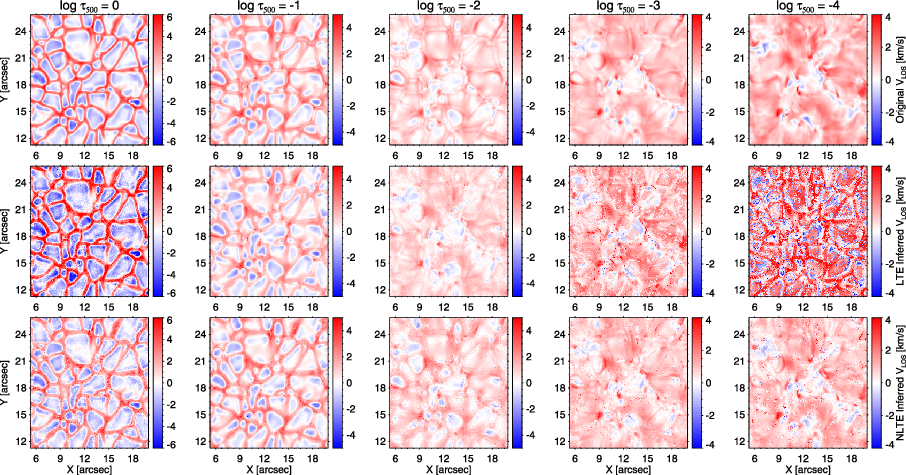}
		\vspace{-0.10cm}
		\caption{Same as Figure~\ref{Diff_T} for the LOS velocity.}
		\label{Diff_Vlos}
	\end{center}
\end{figure*}

\begin{figure*}
	\begin{center} 
		\includegraphics[trim=0 0 00 0,width=18cm]{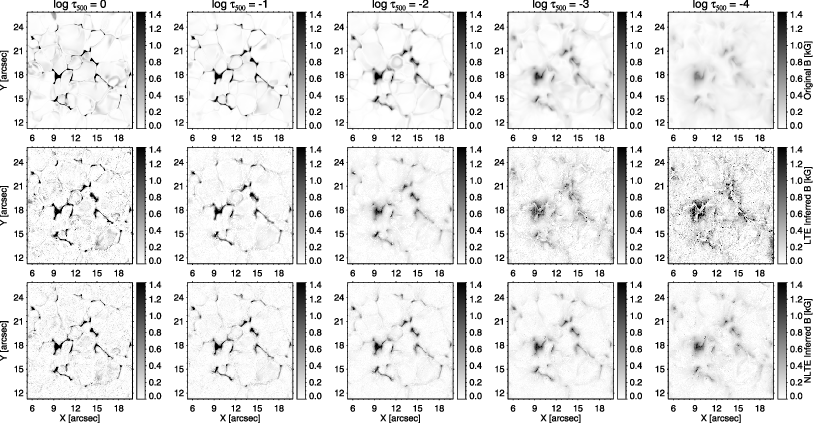}
		\vspace{-0.10cm}
		\caption{Same as Figure~\ref{Diff_T} for the magnetic field strength.}
		\label{Diff_B}
	\end{center}
\end{figure*}

\begin{figure*}
	\begin{center} 
		\includegraphics[trim=0 0 00 0,width=18cm]{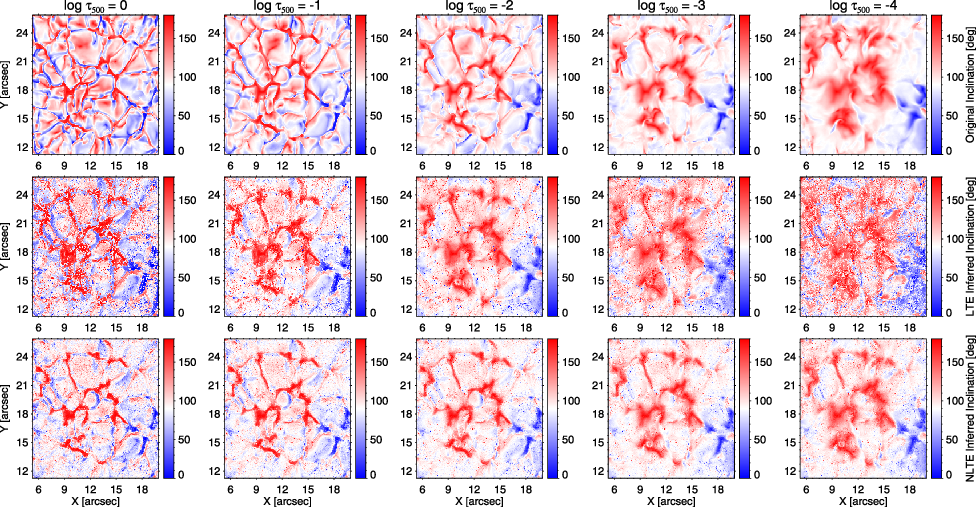}
		\vspace{-0.10cm}
		\caption{Same as Figure~\ref{Diff_T} for the magnetic field inclination.}
		\label{Diff_Incli}
	\end{center}
\end{figure*}

\begin{figure*}
	\begin{center} 
		\includegraphics[trim=0 0 00 0,width=18cm]{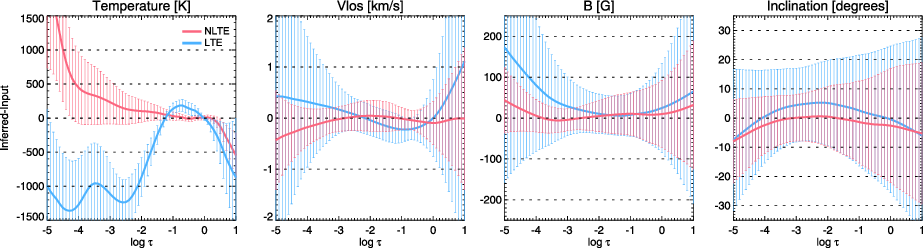}
		\vspace{-0.10cm}
		\caption{Average differences between the input and the inferred atmosphere at different optical depths. From left to right, we show the differences in temperature, LOS velocity, magnetic field strength, and inclination. Colours correspond to the LTE case (blue) and NLTE assumption (red). Error bars display the standard deviation of the difference over the selected FOV at each optical depth.}
		\label{Diff}
	\end{center}
\end{figure*}

Regarding the LOS velocity, we detect a similar behaviour (see Figure~\ref{Diff_Vlos}). Results at around $\log \tau=0$ are similar between LTE and NLTE, although the former shows systematically larger velocities. In this case, results at $\log \tau=-1$ are also good for both inversion schemes. Substantial deviations appear at $\log \tau=-2$, with the NLTE inversion closely resembling the input atmosphere up to $\log \tau=-4$ while the LTE solutions are less accurate. A similar behaviour is found for the magnetic field strength (Figure~\ref{Diff_B}) and inclination (Figure~\ref{Diff_Incli}). These results mean that the LTE inversion produces atmospheres that resemble the input simulation only at the lower atmosphere, below $\log \tau=-2$. In contrast, the NLTE inversion has better accuracy up to $\log \tau\sim-4$.

In Figure~\ref{Diff}, we show the average difference between the input atmosphere and that inferred, assuming LTE (blue) or NLTE (red). Except in some layers at the lower photosphere, the average difference for the LTE inversion is always worse than that from the NLTE inversions. Additionally, for the NLTE inversions, as predicted by the response function analysis shown in previous sections, the differences for temperature become large above $\log \tau=-2$. However, the differences remain small up to $\log \tau\sim-3$ for the rest of the atmospheric parameters.

Finally, in terms of computational time, the average inversion time (just for a single atmospheric model) is $3.26\pm0.56$~s in LTE and $33.99\pm6.03$~s in NLTE. There is roughly an order of magnitude difference in computing time between the two approximations. In any case, with enough CPU cores, the inversion can be carried out in a relatively short time, allowing us to invert the selected FOV in a few hours. Moreover, it will be worth exploring solutions based on machine learning to accelerate the NLTE inversions \citep[e.g.,][]{2022ApJ...928..101V,2022A&A...658A.182C}.

\section{Summary}

This work aimed to evaluate how much can be gained by performing NLTE inversions of the \ion{Si}{i} 10827~\AA \ spectral line using the simplest atom possible. The RFs indicate that the sensitivity of the NLTE solution extends to the upper photospheric layers as we correctly model the line core wavelengths of the transition in contrast with the LTE case. Interestingly, we also see that the NLTE inferred atmosphere remains accurate up to $\log \tau\sim-3$ and slightly higher. In contrast, the LTE atmospheres are correct only at the lower photosphere, below $\log \tau=-2$, as this approximation is only valid for the wings of the spectral line. Also noteworthy is that the computational time increases from an average of 3 s in the LTE case to around 30 s for the NLTE inversions, a factor 10 that, although sizeable, is still smaller than the time needed to invert other NLTE spectral lines like the \ion{Ca}{ii} infrared transitions. Consequently, its impact when inverting simultaneously several photospheric and chromospheric lines is much reduced. Finally, it is also critical that we see no effect on the robustness of the inversion, i.e., no convergence issues were found in the NLTE inversion. Therefore, we conclude that the simplified silicon atom presented in \cite{Shchukina2017} is excellent for tackling the inversions of the \ion{Si}{i} 10827~\AA \ spectral line. 
 
Finally, we plan to use this atom in the upcoming simultaneous observations of the spectral windows at 8542 and 10830~\AA \ at the upgraded GRIS. This instrument will be offered publicly from the second semester of 2024. We believe the \ion{Si}{i} 10827~\AA \ spectral line is an excellent photospheric complement to the already capable \ion{Ca}{ii} 8542~\AA \  transition, promising to increase the height resolution of the inferred atmospheric parameters and the sensitivity to quiet Sun magnetic fields in the lower atmosphere.

\begin{acknowledgements}
C. Quintero Noda, A. Asensio Ramos, T. del Pino Alemán, M. J. Martínez González, and J. C. Trelles Arjona acknowledge support from the Agencia Estatal de Investigación del Ministerio de Ciencia, Innovación y Universidades (MCIU/AEI) under grant ``Polarimetric Inference of Magnetic Fields'' and the European Regional Development Fund (ERDF) with reference PID2022-136563NB-I00/10.13039/501100011033. The publication is part of the Project ICTS2022-007828, funded by MICIN and the European Union NextGenerationEU/RTRP. 
\end{acknowledgements}

\bibliographystyle{aa} 
\bibliography{silicon} 

\end{document}